\documentclass[aps,pra,twocolumn,nopacs,superscriptaddress,floatfix]{revtex4}

\usepackage{amsmath} \usepackage{bbm}
\usepackage{epsfig}

\newcommand{\id}{\mathbbm{1}}

\usepackage{times}

\usepackage{amssymb}
\usepackage{amsmath}
\usepackage{color}

\begin{document}

\title{A proposed testbed for detector tomography}

\author{H.B.\ Coldenstrodt-Ronge,
J.S.\ Lundeen}

\affiliation{Clarendon Laboratory, University of Oxford, Oxford,
OX1 3PU, UK}

\author{K.L.\ Pregnell}
\author{A.\ Feito}

\affiliation{Institute for Mathematical Sciences, Imperial College
London, London SW7 2PG, UK} 
\affiliation{QOLS, Blackett Laboratory, Imperial
College, London SW7 2BW, UK}

\author{B.J.\ Smith}

\affiliation{Clarendon Laboratory, University of Oxford, Oxford,
OX1 3PU, UK}

\author{W.\ Mauerer, Ch.\ Silberhorn},

\affiliation{Max-Planck Research Group,
Guenther-Scharowskystrasse 1/Bau 24, 91058 Erlangen, Germany}
 
\author{J.\ Eisert}

\affiliation{QOLS, Blackett Laboratory, Imperial
College, London SW7 2BW, UK}
\affiliation{Institute for Physics and Astronomy, University of
Potsdam, 14476 Potsdam, Germany}

\author{M.B.\ Plenio}

\affiliation{Institute for Mathematical Sciences, Imperial College
London, London SW7 2PG, UK} 
\affiliation{QOLS, Blackett Laboratory, Imperial
College, London SW7 2BW, UK}

\author{I.A.\ Walmsley}

\affiliation{Clarendon Laboratory, University of Oxford, Oxford,
OX1 3PU, UK}

\begin{abstract}
Measurement is the only part of a general quantum system that has yet to be
characterized experimentally in a complete manner. Detector tomography
provides a procedure for doing just this; an arbitrary measurement device
can be fully characterized, and thus calibrated, in a systematic way without
access to its components or its design. The result is a reconstructed POVM
containing the measurement operators associated with each measurement
outcome. We consider two detectors, a single-photon detector and a
photon-number counter, and propose an 
easily realized experimental apparatus to perform detector tomography on them. We also present a 
method of visualizing the resulting
measurement operators.
\end{abstract}

\maketitle

\section{Introduction}

A quantum protocol or experiment can be divided into three stages:
preparation, processing, and measurement. Quantum state tomography \cite%
{Smithey1993,Dunn1995,White1999} and process tomography \cite%
{Childs2001,Mitchell2003} respectively prescribe a procedure to completely
characterize the first two stages, and have been successfully demonstrated
experimentally. State tomography has played a crucial role in identifying
and visualizing novel quantum states \cite{White2001, Lvovsky2001}. Process
tomography is critical for verifying the operation of quantum logic gates 
\cite{O'Brien2004} and characterizing decoherence processes \cite%
{Myrskog2005}. Completing this triad, detector tomography is a procedure for
determining the POVM (positive operator-valued measure) set of an arbitrary
quantum detector \cite{Luis1999}. Thus, without any knowledge of the inner
workings of the detector we could predict its response to any input. To the
best of our knowledge, detector tomography has not been demonstrated. Here, we
propose an experimental testbed capable of performing detector
tomography on measurement devices acting in the Fock space of the optical mode (i.e., the photon number Hilbert space).
Precise knowledge of the detection POVM set is beneficial for measurement
driven quantum information processing, such as cluster-state computing \cite%
{Raussendorf2001}. It is also critical for state and process tomography,
which use the measurement as a reference. Recently, it has also been shown
that one can perform enhanced measurements by projecting onto non-classical
states with a detector \cite{Resch2007, Resch2007a}. Without any need for a
theoretical model of the detector, detector tomography can establish which
states the detector projects onto, possibly establishing whether they are
indeed non-classical.

We will begin by introducing the general experimental and theoretical
requirements for performing detector tomography. We will then describe the
detectors which we aim to characterize using our proposed testbed. Following this, we
will provide a description of the testbed we have proposed to characterize
the aforementioned detectors. Finally, we outline a method of visualizing
the POVM elements in terms of Wigner quasi-probability distributions.

\section{Detector Tomography}

Detector tomography is a method of experimentally determining the POVM
associated with the detector. A theoretical description of detector
tomography was introduced by Luis and S\'{a}nchez-Soto in Ref.\ \cite%
{Luis1999} in 1999, a maximum-likelihood technique was applied in Ref.\ 
\cite{Fiurasek2001} in 2001, ancilla-assisted detector tomography was
considered in 2004 in Ref.\ 
\cite{D'Ariano2004}, and an optimal processing scheme
devised in 2007 in Ref.\ 
\cite{D'Ariano2007}.

In elementary quantum mechanics, a measurement is described by a Hermitian
operator $\hat{A}$ which can be decomposed into a sum of projectors $\hat{\pi%
}_{i}$ with weights $\lambda_{i}$, where each projector corresponds to an
outcome of the measurement: 
\begin{equation}
\hat{A}=\sum_{i}\lambda_{i}\hat{\pi}_{i}.
\end{equation}
These projectors $\hat{\pi}_{i}$ are often deduced by the action of the
measuring device in the classical regime. For example, since a calcite
crystal spatially separates the polarization components of a laser beam it
is reasoned to act similarly at the quantum level on single photons (i.e., $%
\hat{\pi}_{1}=\left\vert H\right\rangle \left\langle H\right\vert$ and $\hat{%
\pi}_{2}=\left\vert V\right\rangle \left\langle V\right\vert$, horizontal
and vertical polarization projectors aligned with the axes of the calcite).
This example is justified by a theoretical model, namely
quantum-electrodynamics, which shows the classical field reduces to field
operators at the quantum level. However, some measurements, such as
Bell-state measurement \cite{Mitchell2003} or single-photon detection,
cannot be traced back to any classical device. These \textit{quantum}
detectors typically rely on an unsystematic combination of a theoretical
model (e.g., semiconductor theory) and detector characterizations of limited
scope (e.g., of the detector noise and efficiency). In contrast, detector
tomography would provide a systematic method, with minimal assumptions, to
characterize these quantum detectors and, thus, predict their action on an
general input state.

The most general type of quantum measurement is described by a POVM, a set
of positive-semidefinite operators $O_{\alpha }$ corresponding to the
measurement outcomes. These generalized measurements are uniquely quantum in
the sense that they have no analogue in the classical regime. For instance,
any POVM\ can be implemented by coupling the measured system to an ancilla
system and then performing projective measurements on the combined system 
\cite{Peres1990}; the coupling has the potential to create entanglement
between the two systems, which is impossible in the classical regime.

In some ways, detector tomography is very similar to the established
procedure of state tomography. In state tomography, one begins with an
ensemble of systems prepared identically in state $\hat{\rho}$. A
measurement $\hat{O}_{\alpha }$, is performed on a subset of the ensemble
for the purpose of estimating the probability $p_{\alpha }$: 
\begin{equation}
p_{\alpha }=\mathrm{Tr}(\hat{O}_{\alpha }\hat{\rho}).  \label{trace}
\end{equation}%
This is repeated for a set of 
measurements $\{\hat{O}_{\alpha }\}$,
producing a set of 
estimated probabilities $\{p_{_{\alpha }}\}$. Through
Eq.\ (\ref{trace}) and the precise knowledge of the form of each
measurement $\hat{O}_{\alpha }$, one can estimate $\hat{\rho}$.
Significantly, the role of $\hat{O}_{\alpha }$ and $\hat{\rho}$ are
symmetric in Eq.\ (\ref{trace}). 
This means that with a set of known input
states $\{\hat{\rho}_{\alpha }\}$ one could instead estimate an unknown measurement $\hat{O}$ through Eq.\ (\ref{trace}),
which is the goal of detector tomography.

Despite these similarities, state and detector tomography have some
important differences. One obvious difference is that the former seeks to
reconstruct a single operator, $\hat{\rho}$, whereas the latter seeks to
reconstruct a set of operators $\{\hat{O}_{\gamma }\},\gamma =1,\ldots ,D$,
where $D$ is the number of measurement outcomes. Both the density matrices
and the POVM elements must be positive-semidefinite and, hence, Hermitian.
However, density matrices have a 
unit trace, whereas a POVM element does
not. Instead, a POVM\ set $\{\hat{O}_{\gamma }\}$ must satisfy, 
\begin{equation}
\sum_{\gamma =1}^{D}\hat{O}_{\gamma }=\hat{\id},  \label{sumtoone}
\end{equation}%
where $\hat{\id}$ is the identity, ensuring that the probabilities for all the
measurement outcomes sum to one. This has implications for mathematical
strategies for reconstruction, such as maximum likelihood, where Eqs.\ %
(\ref{trace}) and (\ref{sumtoone}) 
must be included as constraints on the
reconstructions.

Generally, in tomography one requires what is known as a \textquotedblleft
tomographically complete\textquotedblright\ set to reference against. In
state tomography, this translates to a set of reference measurements $\{\hat{%
O}_{\alpha }\}$ that span the 
$d^{2}-1$ dimensional 
Hilbert-Schmidt space of the density matrix
(the $-1$ term comes from the unit trace
constraint). In detector tomography, 
the reference states $\{\hat{\rho}%
_{\alpha }\}$ must span the Hilbert-Schmidt space of the POVM set. A
spanning set will necessarily have at least $d^{2}$ elements in it. To
determine if a particular $d^{2}$ sized subset of the total set spans the
space, one first vectorizes each element in the subset, then stacks these
row vectors into a matrix, and then calculates the determinant. If the
determinant is nonzero this subset is a spanning set and hence, the set of
reference states (or reference measurements, in the case of state
tomography) is tomographically complete. We shall consider an example later
in the paper.

\section{The detectors}

Before performing detector tomography, one needs to analyze the detectors in
order to identify the Hilbert space they function in, and then find a
suitable set of input states. We propose a testbed for two types of
related detectors: the avalanche photodiode (APD), and the time-multiplexing
detector (TMD).

APDs are photodiodes that use the avalanche effect (i.e., impact ionization
leading to the exponential multiplication of carriers) to achieve
sensitivity to single photons. These detectors have the remarkable ability
to detect a single photon. Unfortunately, if more than one photon is
detected this information is lost in the avalanche. Hence, this detector
acts as a binary detector; the two possible detection outcomes are detection
of at least one photon (\textsc{click}) and the detection of no photons (%
\textsc{no click}). For each photon impinging on the detector there is an
intrinsic efficiency of the detector $\eta _{\text{APD}}$ that the photon
will cause an avalanche. The probability 
of detecting at least one out of $n$
photons is given by 
\begin{equation}
P(\text{\textsc{click}})_{n}=1-\left( 1-\eta _{\text{APD}}\right) ^{n},
\label{pclick}\end{equation}%
and the probability of detecting none of the $n$ 
photons is, 
\begin{equation}
P(\text{\textsc{no click}})_{n}=\left( 1-\eta _{\text{APD}}\right) ^{n}.
\label{pnclick}\end{equation}%
Thus, there is a non-zero probability of getting a \textsc{no click} event,
even with photons hitting the detector. The detection can be approximated
with the two POVM elements, 
\begin{eqnarray}
\text{\textsc{no click}} &\emph{:}&\hat{O}_{0}=\sum_{n=0}^{\infty }\left(
1-\eta _{\text{APD}}\right) ^{n}\left\vert n\right\rangle \left\langle
n\right\vert , \\
\text{\textsc{click}} &:&\hat{O}_{1}=\hat{\id}-\sum_{n=0}^{\infty }\left(
1-\eta _{\text{APD}}\right) ^{n}\left\vert n\right\rangle \left\langle
n\right\vert .
\end{eqnarray}%
These approximate POVM elements do not include effects such as dark counts, afterpulsing, and detector saturation. The contributions from dark counts and afterpulsing can cause detection events without an actual photon being present. They will increase the values of the $\hat{O}_{1}$ matrix elements. However since dark counts are independent of the actual count rate they should predominantly affect the matrix elements at low photon numbers. Conversely, afterpulsing is dependent on a counting event happening before, and thus, would be more prominent for higher photon numbers. Detector saturation would manifest itself by a dependence of $\eta _{\text{APD}}$ on $n$. These effects would have to be included in a more complicated detector model if one desired to derive more realistic POVM elements. Previous experiments with APDs \cite{Coldenstrodt2007} indicate that these approximate POVM elements will be a reasonable model on which to design a testbed.

\begin{figure}[tbp]
\centering{\ \includegraphics[width=0.5\textwidth]{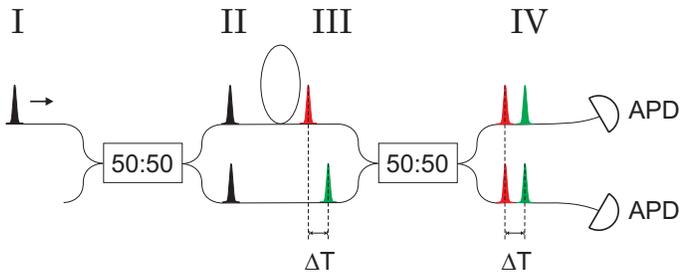} }
\caption{Schematic of a time multiplexed detector. The incoming pulse ($%
\mathbf{I}$) is split at a 50:50 beam splitter. The resulting pulses ($%
\mathbf{II}$) are partially delayed ($\mathbf{III}$) and then split again ($%
\mathbf{IV}$). The initial input pulse ($\mathbf{I}$) ends up in several
temporal and spatial bins ($\mathbf{IV}$), which are then detected with
APDs. }
\label{loopyschematics}
\end{figure}

Several schemes have been proposed to overcome the lack of photon number
resolution with APDs \cite{Paul1996,Kok2001,Banaszek2003}. In a time
multiplexing detector, the pulse under investigation is split into several
spatial and temporal bins and then detected with APDs \cite%
{Achilles2003,Coldenstrodt2007,Fitch2003}. The principle of operation of our
implementation is depicted in Figure \ref{loopyschematics}. 

The pulse under investigation ($\mathbf{I}$) is split at a beamsplitter and
one of the two spatial modes ($\mathbf{II}$) is delayed with respect to the
other one ($\mathbf{III}$). The two spatial modes are then recombined at a
beamsplitter, resulting in two temporal modes in each of the spatial outputs
of the beamsplitter ($\mathbf{IV}$). Additional delay stages can be added to
further increase the number of temporal modes. We propose to perform detector
tomography on a TMD with two delay stages, resulting in a total of eight
output bins. The advantage of splitting the incoming pulse into many bins is
that if the pulse contains several photons these may be distributed into
different bins and, thus, counted as individual photons by APDs. The
measurement result of the TMD is the number of \textsc{click} events summed
over all the bins. Thus, for our detector, with two stages, there are nine
outcomes, 0 to 8 \textsc{clicks}. Unfortunately, $k$ \textsc{clicks} does
not imply there were only $k$ photons at the TMD input. Photons may end up
in the same bin and, thus, result in fewer clicks than the number of photons
in the input pulse. Photons also suffer from losses in the TMD\ and the
detection efficiency of the APDs. While this introduces an uncertainty in
the measurement outcome on a shot-to-shot basis, for an ensemble measurement
of the same quantum state the resulting click-statistics 
grasped in the vector ${p}$ can be
related to the photon number statistics 
as a vector ${\sigma}$ of the input state by 
\begin{equation}
	{p}=\mathbf{C}\cdot \mathbf{L}{\sigma}, \label{loopyequation}
\end{equation}%
where $\mathbf{L}$ represents the losses in the
system and $\mathbf{C}$ is called the convolution matrix, which accounts for
several photons ending up in the same bin after splitting. The loss matrix
can be calculated by combining all losses in the system and modeling them
as a single beam splitter with reflectivity $\eta _{\text{loss}}$ at the
front of the fiber network \cite{Silberhorn2004,Achilles2006} as, 
\begin{equation}
\mathbf{L}_{n\prime n}=%
\begin{cases}
\binom{n}{n\prime }\left( 1-\eta _{\text{loss}}\right) ^{n\prime }\eta _{%
\text{loss}}^{n-n\prime } & \text{if:\ }n\geqslant n\prime  \\ 
0 & \text{otherwise,}%
\end{cases}%
\end{equation}%
with $n,n\prime \in 0,\dots, N$. This simply
describes the probability of $n\prime $ out of $n$ photons being
transmitted. For the convolution matrix one has to take into account all
possible routes a photon could take through the fiber network \cite%
{Achilles2004} and the matrix elements can be calculated as follows: 
\begin{widetext}
\begin{equation}
\mathbf{C}_{k,n}=%
\begin{cases}
0, & k>n, \\ 
\sum_{b}p_{b_{1}}p_{b_{2}}\cdots p_{b_{n}}, & k=n, \\ 
\sum_{b}\left[ \sum_{c}\frac{1}{{\operatorname{id}(c)!}}\prod_{j=1}^{k}\binom{%
n-\sum_{l=0}^{j-1}c_{l}}{c_{j}}%
p_{b_{1}}^{c_{1}}p_{b_{2}}^{c_{2}}
\cdots 
p_{b_{k}}^{c_{k}}\right],  & k<n.%
\end{cases}%
\label{fancyC}\end{equation}
\end{widetext}

In reality, the beamsplitters used in the TMD are never exactly 50\%
reflective due to variation in their manufacture. For a $N$-bin TMD, $%
p_{1},\ldots ,p_{N}$ are the probabilities of a single photon exiting from the
fiber network in a particular bin. While these probabilities would all be
equal in the case of 50/50 beamsplitters, we set them by calibration
measurements using an intense laser input pulse. The first two lines of
Eq.\ (\ref{fancyC}) 
correspond to the straightforward cases of more photons
being detected than entering the detector (for which the probability is zero,
if dark counts and afterpulsing are negligible) and detecting all incoming
photons respectively. In the third case, $k<n$, all possible combinations of
distributing these $n$ photons into the bins have to be taken into
account. These different combinations are represented by the $k$-tuples $%
b=\left( b_{1},b_{2},\cdots,b_{k}\right) $ with $b_{i}\in 
1,\dots, N$ 
and $b_{1}\neq b_{2}\neq \cdots\neq b_{k}$. Some of the bins can be occupied by more than one photon. All possible
distributions of the photons across the bins are described by the possible
partitions of \(n\) into \(k\) parts.  These 
can be represented by $k$-tuples
$c=\left( c_{1},c_{2},\cdots,c_{k}\right) $ with $c_{i}\in
\{1,\dots, n\}, \text{and }c_{1}\geqslant c_{2}\geqslant \cdots\geqslant c_{k}$,
\(\sum_{k}c_{k} = n\), and the definition $c_{0}=0$. For each bin, one has to consider all possible ways in which the $c_j$ photons ending up in this particular bin can be choosen. With the photons remaining to be distributed given by $n-\sum_{l=0}^{j-1}c_{l}$, the binomial coefficient in Eq.\ (\ref{fancyC}) accounts for this. However, if some of the elements of the tuple $c$ are equal, the binomial coefficient will overestimate the number of distributions.
Bins occupied with the same number of photons are not distinguishable, but
are counted separately. This must be corrected
for by dividing by the number of permutations these bins can form, i.e., the
factorial of the number of bins with equal number of photons. We denote
this by $\operatorname{id}(c)!$. For example for the tuple $c=\left(4;4;2;2;2;1\right)$, $\operatorname{id}(c)!=2!\cdot3!$.

For the eight bin TMD that we plan to use in the proposed detector tomography scheme, the
convolution matrix was calculated using classical measurements of the
probabilities $p_{1},\ldots ,p_{N}$ and reads: 
\begin{equation}
\mathbf{C}=%
\begin{pmatrix}
1 & 0 & 0 & 0 & 0 & 0 & 0 & 0 & 0 \\ 
0 & 1 & 0.128 & 0.017 & 0.000 & 0.000 & 0.000 & 0.000 & 0.000 \\ 
0 & 0 & 0.872 & 0.334 & 0.101 & 0.028 & 0.008 & 0.002 & 0.001 \\ 
0 & 0 & 0 & 0.649 & 0.496 & 0.265 & 0.123 & 0.053 & 0.022 \\ 
0 & 0 & 0 & 0 & 0.402 & 0.509 & 0.422 & 0.290 & 0.181 \\ 
0 & 0 & 0 & 0 & 0 & 0.198 & 0.375 & 0.444 & 0.423 \\ 
0 & 0 & 0 & 0 & 0 & 0 & 0.073 & 0.193 & 0.308 \\ 
0 & 0 & 0 & 0 & 0 & 0 & 0 & 0.018 & 0.063 \\ 
0 & 0 & 0 & 0 & 0 & 0 & 0 & 0 & 0.002 \\ 
&  &  &  &  &  &  &  & 
\end{pmatrix}%
.\label{cmsmtmd}
\end{equation}%

The matrices developed above actually contain the POVM\ elements describing
the TMD. We expect these POVM\ elements to be diagonal in the photon number
basis since the TMD has no phase reference and, thus, no sensitivity to
off-diagonal coherences. In Eq.\ 
(\ref{loopyequation}), the matrix $%
\mathbf{C}\cdot \mathbf{L}$ represents the reaction of the detector to
different numbers of incoming photons. When the matrix $\mathbf{CL}$ acts on
a photon number distribution ${\sigma}$, the probability of a particular
measurement outcome is found by taking an inner product of ${\sigma}$
with the corresponding row of the matrix. This action is similar to the
action of the trace in Eq.\ (\ref{trace}), as we will demonstrate
explicitly. This allows us to identify the rows of $\mathbf{C}\cdot \mathbf{L%
}$ as the diagonals of the TMD's POVM: 
\begin{equation}
\hat{O}_{j}=%
\begin{bmatrix}
\left[ \mathbf{C}\cdot \mathbf{L}\right] _{j,0} & 0 & 0 & 0 & 0 \\ 
0 & \ddots  & 0 & 0 & 0 \\ 
0 & 0 & \left[ \mathbf{C}\cdot \mathbf{L}\right] _{j,i} & 0 & 0 \\ 
0 & 0 & 0 & \ddots  & 0 \\ 
0 & 0 & 0 & 0 & \left[ \mathbf{C}\cdot \mathbf{L}\right] _{j,N}%
\end{bmatrix}%
,
\label{POVMwithCL}\end{equation}%
 where $j$ is the $j$-\textsc{clicks} outcome, and $\left[ 
\mathbf{C}\cdot \mathbf{L}\right] _{j,i}$ is the value in the $j$th row and $i
$th column of $\mathbf{C}\cdot \mathbf{L}$. The density matrix of the
incoming quantum state is related to ${\sigma}$ via, 
\begin{equation}
\hat{\rho}=%
\begin{bmatrix}
\sigma _{0} & \rho _{0,1} & \cdots  & \rho _{0,N} \\ 
\rho _{1,0} & \sigma _{1} & \cdots  & \rho _{1,N} \\ 
\vdots  & \vdots  & \ddots  & \vdots  \\ 
\rho _{N,0} & \rho _{N,1} & \cdots  & \sigma _{N}%
\end{bmatrix}%
,
\end{equation}%
expressed in the number basis, where the $i$th element of 
${\sigma}$ is $%
\sigma _{i}$. We leave the off-diagonal elements of $\hat{\rho}$ as
undetermined, since they do not contribute after the trace. Using this form
we evaluate the trace in Eq.\ (\ref{trace}): 
\begin{widetext}
\begin{align}
\mathrm{Tr}(\hat{O}_{j}\hat{\rho})& =\mathrm{Tr}\left( 
\begin{bmatrix}
\left[ \mathbf{C}\cdot \mathbf{L}\right] _{j,0} & 0 & 0 & 0 & 0 \\ 
0 & \ddots  & 0 & 0 & 0 \\ 
0 & 0 & \left[ \mathbf{C}\cdot \mathbf{L}\right] _{j,i} & 0 & 0 \\ 
0 & 0 & 0 & \ddots  & 0 \\ 
0 & 0 & 0 & 0 & \left[ \mathbf{C}\cdot \mathbf{L}\right] _{j,N}%
\end{bmatrix}%
\cdot 
\begin{bmatrix}
\sigma _{0} & \rho _{0,1} & \cdots  & \rho _{0,N} \\ 
\rho _{1,0} & \sigma _{1} & \cdots  & \rho _{1,N} \\ 
\vdots  & \vdots  & \ddots  & \vdots  \\ 
\rho _{N,0} & \rho _{N,1} & \cdots  & \sigma _{N}%
\end{bmatrix}%
\right)  \\
& =\mathrm{Tr}\left( 
\begin{bmatrix}
\left[ \mathbf{C}\cdot \mathbf{L}\right] _{j,0}\cdot \sigma _{0} & 0 & 0 & 0
& 0 \\ 
0 & \ddots  & 0 & 0 & 0 \\ 
0 & 0 & \left[ \mathbf{C}\cdot \mathbf{L}\right] _{j,i}\cdot \sigma _{i} & 0
& 0 \\ 
0 & 0 & 0 & \ddots  & 0 \\ 
0 & 0 & 0 & 0 & \left[ \mathbf{C}\cdot \mathbf{L}\right] _{j,N}\cdot \sigma
_{N}%
\end{bmatrix}%
\right)  \\
& =\sum_{i=0}^{N}\left[ \mathbf{C}\cdot \mathbf{L}\right] _{j,i}\cdot \sigma
_{i}=p_{j}.
\end{align}%
This last line is equivalent to Eq.\ (\ref{loopyequation}), as we
expected.
\end{widetext}

As we have shown, one can derive a theoretical POVM set to describe a
time-multiplexed detector. However, as with the APD, this is a simplified
description of the POVM and does not include afterpulsing (which can result
in a click from one time bin triggering a click in the next), dark counts in
the APDs at the outputs, as well as imperfections in the counting
electronics (which can be considerably complicated).

In summary, the APD and the TMD perform their measurements in the Fock space of the optical mode. Furthermore, both detectors have POVMs that are expected to be diagonal in the photon-number basis. Since this confines their action to a subspace of the Fock space, it should allow us to simplify the tomography. For the APDs all photon numbers apart from zero
photons are combined into a single detection event and, to a first
approximation, the only detector parameter is the efficiency. This makes the
APD a rather simple detector and a theoretical model of their POVM elements
was easily calculated. Thus, APDs would be the first detector with which we propose to demonstrate a proof of principle detector tomography. The TMD, on
the other hand, is non-trivial to describe, with both the loss and the
convolution matrix having complicated forms. Incorporating realistic APDs
into this model would be challenging, if not impossible. Consequently, the
TMD is the second detector that we propose to perform detector tomography on. It is a
suitable candidate for a serious test of the detector tomography procedure.
We propose to compare the reconstructed POVM set to the theoretical ones
derived above to check how accurate our detector models really are.

\section{The proposed testbed}

To perform tomography of these two detectors, we need probe states in the Fock space. We must also identify a set of states that span
the space of the detector operators, i.e., a tomographically complete set. Since neither
detector possesses a phase-reference (such as a local oscillator), we make
the safe assumption for our proposed testbed that the off-diagonal components of $\{%
\hat{O}_{\gamma }\}$ for each detector are zero in the photon-number basis.
It follows that each $\hat{O}_{\gamma }$ contains $d$ free parameters, one
for each diagonal element, where $d$ is the the dimension of the Hilbert
space. The tomographically complete set will then contain at least $d$ input
states. Unfortunately, the photon number representation of the Hilbert space is infinite, and so
we must truncate it at some point for our mathematical reconstruction. This
truncation sets the $d$ we shall use in the tomography testbed we propose. A good point
to set this truncation at is in the region in which the detector behavior
has saturated to a constant. e.g. We expect the TMD's response to 100
photons and 101 photons will be nearly identical (outcome 8-\textsc{clicks}
with $\approx 100\%$ probability), making $d=100$ a good place to truncate.
Since we do not require sensitivity to inter-photon coherences a naturally
tomographically complete set of input states $\{\hat{\rho}_{\alpha }\}$ for
our proposed testbed is the Fock states. Each input Fock state would allow us to
determine the corresponding diagonal element in each of the TMD's
measurement operators $\{\hat{O}_{\gamma }\}.$ However, Fock states are
difficult to produce with high fidelity. Fortunately, another candidate, the
coherent state, is straightforward to produce with high-fidelity.

Apart from being the quantum state that most resembles a classical state,
coherent states have a unique property under attenuation: in the
photon-number basis, a coherent state $|\alpha\rangle $ can be written, 
\begin{equation}
	|\alpha\rangle=\sum_{n=0}^{\infty }
	{{\exp}\left( {-\frac{|\alpha |^{2}}{2}%
}\right) \frac{\alpha ^{n}}{\sqrt{n!}}|n\rangle,}  \label{equ_poisson}
\end{equation}%
which results in a Poissonian photon-number distribution. With attenuation $%
1-\eta $, the coherent state transforms according to, 
\begin{equation}
	\hat{U}_{\eta }|\alpha\rangle=|\eta\alpha\rangle
=\sum_{n=0}^{\infty }{{\exp%
}\left( {-\frac{\eta |\alpha |^{2}}{2}}\right) \frac{\left( \sqrt{\eta }%
\alpha \right) ^{n}}{\sqrt{n!}}|n\rangle.}  \label{equ_attenuation}
\end{equation}%
This is another coherent state with a lower mean number of photons. This
property is important for the creation of known states in the low photon
number regime, a necessity for tomography on single photon detectors.
Specifically, the high-fidelity coherent states produced by a laser could be
easily attenuated to power levels suitable for the tomography while still
retaining their form. Attenuation also reduces the impact of technical noise
in the laser. The fractional uncertainty in the photon number of a coherent
state, ${\Delta {n}}/{\langle n\rangle}$, is - once it is attenuated to a low photon
level - much higher than the pulse-to-pulse jitter of the pulse energy of
our laser source (typical values around $2\%$). This suggests that the use
of coherent states would also be reasonably robust against technical noise.

It is well known, that a continuous set of coherent probe 
state vectors 
$|\alpha\rangle $ form a basis for the
Fock space, 
\begin{equation}
|\psi \rangle =\int d\alpha |\alpha \rangle\langle\alpha |\psi \rangle.
\end{equation}%
However it is not as clear that they form a basis for the Hilbert-Schmidt space. The proof that they do lies in the existence of the P-function $P(\alpha)$:
\begin{equation}
\hat{\rho}=\int P(\alpha) |\alpha \rangle\langle\alpha| d^2\alpha.
\end{equation}%
However, we expect to use a discrete subset of coherent states $|\alpha\rangle_{i}$ rather than the full continuous basis for our proposed detector tomography testbed. We now attempt to find if this reduced set is tomographically complete. We
vectorize $d$ coherent states density matrices $\left\vert \alpha
_{i}\right\rangle \left\langle \alpha _{i}\right\vert ,$ $i=1,\ldots ,d$,
keeping only the diagonals (since our detector POVM set is contained within
this subspace). Stacking these row vectors into a matrix, we find, 
\begin{equation}
\begin{bmatrix}
\left\vert \left\langle 0|\alpha _{1}\right\rangle \right\vert ^{2} & 
\left\vert \left\langle 1|\alpha _{1}\right\rangle \right\vert ^{2} & \ldots 
& \left\vert \left\langle d|\alpha _{1}\right\rangle \right\vert ^{2} \\ 
\left\vert \left\langle 0|\alpha _{i}\right\rangle \right\vert ^{2} & \ddots 
&  &  \\ 
\vdots  &  & \ddots  &  \\ 
\left\vert \left\langle 0|\alpha _{d}\right\rangle \right\vert ^{2} &  &  & 
\left\vert \left\langle d|\alpha _{d}\right\rangle \right\vert ^{2}%
\end{bmatrix}%
.
\end{equation}%
The next step, finding the determinant, is difficult to do in general.
Instead we consider a specific set of $d=10$ reference coherent states, in
evenly spaced steps from $|\alpha _{1}|^{2}=1$ to $|\alpha _{10}|^{2}=10$.
The determinant 
\begin{equation}
{\det}\left( 
\begin{bmatrix}
0.37 & 0.37 & 0.18 & 0.06 & 0.02 & 0.00 & 0.00 & 0.00 & 0.00 & 0.00 \\ 
0.14 & 0.27 & 0.27 & 0.18 & 0.09 & 0.04 & 0.01 & 0.00 & 0.00 & 0.00 \\ 
0.05 & 0.15 & 0.22 & 0.22 & 0.17 & 0.10 & 0.05 & 0.02 & 0.01 & 0.00 \\ 
0.02 & 0.07 & 0.15 & 0.20 & 0.20 & 0.16 & 0.10 & 0.06 & 0.03 & 0.01 \\ 
0.01 & 0.03 & 0.08 & 0.14 & 0.18 & 0.18 & 0.15 & 0.10 & 0.07 & 0.04 \\ 
0.00 & 0.01 & 0.04 & 0.09 & 0.13 & 0.16 & 0.16 & 0.14 & 0.10 & 0.07 \\ 
0.00 & 0.01 & 0.02 & 0.05 & 0.09 & 0.13 & 0.15 & 0.15 & 0.13 & 0.10 \\ 
0.00 & 0.00 & 0.01 & 0.03 & 0.06 & 0.09 & 0.12 & 0.14 & 0.14 & 0.12 \\ 
0.00 & 0.00 & 0.01 & 0.02 & 0.03 & 0.06 & 0.09 & 0.12 & 0.13 & 0.13 \\ 
0.00 & 0.00 & 0.00 & 0.01 & 0.02 & 0.04 & 0.06 & 0.09 & 0.11 & 0.13%
\end{bmatrix}%
\right) \neq 0,
\end{equation}%
implying that this set is tomographically complete, for a space truncated at photon number basis state $n=d$. Consequently, it appears that a finite set of coherent states are a good set of reference states for our proposed detector tomography testbed.

\begin{figure}[tbp]
\centering{\ \includegraphics[width=0.5\textwidth]{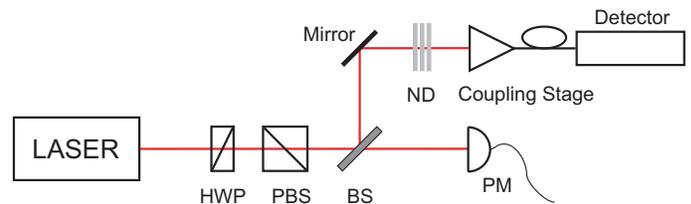} }
\caption{Schematic of our detector testbed. A probe beam undergoes variable
attenuation realized with a half waveplate (HWP) and a polarizing
beamsplitter (PBS). A large portion of the beam is split of at a
beamsplitter (BS) and detected by a power meter (PM) to monitor the variable
attenuation, while the smaller portion of the beam is further attenuated
with neutral density filters (ND) and then coupled into the detector under
investigation.}
\label{optical_setup}
\end{figure}

An essential requirement for a tomography system is precise and accurate
knowledge of the probe states. In the context of our proposed testbed, this
translates to knowing $\alpha $ accurately. Keeping in mind that $|\alpha
|^{2}$ will be less than $100$,
 a direct measurement is out of reach of
commercial power meters. For a coherent state the mean number of photons $%
\langle n\rangle $ is connected to $\alpha $ via, 
\begin{equation}
\langle n\rangle=|\alpha |^{2}.
\end{equation}%
For pulsed light of the wavelength $\lambda $ and the repetition frequency $%
f $ the time averaged power $P$ can be calculated by, 
\begin{equation}
P=\frac{\langle n\rangle hcf}{\lambda },
\end{equation}%
which then enables a calculation of $|\alpha |^{2}$ from the measured power: 
\begin{equation}
|\alpha |^{2}=\frac{P\lambda }{hcf}.
\end{equation}%
While $\lambda$ and $f$ can generally be determined accurately, P cannot. Systematic error in the power measurement would result in a global scaling of the detector response: the reconstructed ``efficiency'' of the detector would scale by the error, while the form of the POVM element would remain unchanged. We expect this $|\alpha|^2$ systematic error to be less than 5\%. For $|\alpha |^{2}=1$, $\lambda =800$ nm and $f=100$ kHz (experimental parameters, we plan to run the proposed tomography on), $P=25$ attoWatts, a strikingly low average power.
However, since coherent states are invariant under attenuation, a highly
transmissive beamsplitter could be used to pick off a large portion of the
incoming beam. If the ratio of power in the transmitted and reflected arms
is well calibrated, the high power arm can be used to monitor the power in
the low power arm. This can leverage the power into the microWatt range,
accessible to power meters with calibrations that can be 
traced to a
National Standards Institute. In our proposed testbed, this beamsplitter would be placed after a variable attenuator, used to set $\alpha $, as depicted in Figure \ref{optical_setup}. For phase-sensitive detectors, a tomographically complete set of input states could be generated simply by adding a variable path length for the coherent state in the testbed, to phase shift
 $\alpha$, i.e., $\alpha\mapsto\alpha e^{i\phi}$.

\begin{figure}[tbp]
\centering{\ \includegraphics[width=0.56\textwidth]{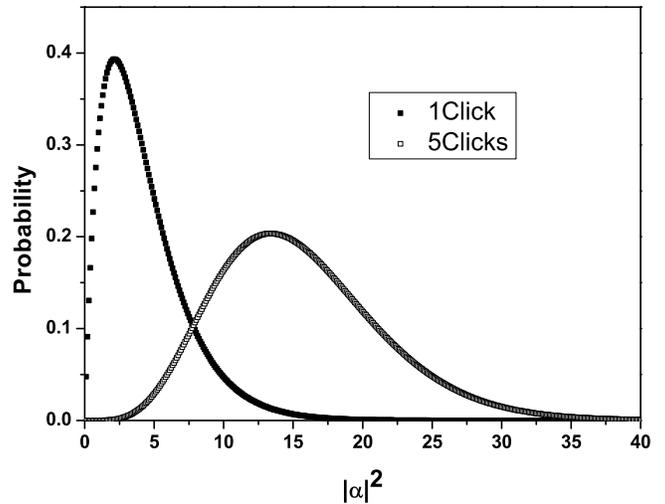} }
\caption{Simulation of a detector tomography for the TMD. 400 different
values of $|\protect\alpha |^{2}$ between $|\protect\alpha |^{2}=0$ and $|%
\protect\alpha |^{2}=40$ were simulated. A comparison between the 1-\textsc{%
click} and the 5\textsc{click} element shows a small overlap and smaller
peak for the 1-\textsc{click} element, while the 5-\textsc{click} element is
broader and peaked at higher photon number.}
\label{simulation}
\end{figure}

We simulated the tomography of a TMD with our proposed testbed by using the model
given in Eq.\ (\ref{loopyequation}). 
We used the convolution matrix for
our TMD displayed in Eq.\ (\ref{cmsmtmd}) 
and simulated a detector tomography using 400
different values of $|\alpha |^{2}$ for the reference states, ranging from $%
|\alpha |^{2}=0$ to $|\alpha |^{2}=40$. In Figure \ref{simulation}, we plot
the probabilities of both the 1-\textsc{click} and 5-\textsc{click} outcomes
against the $|\alpha |^{2}$ of the probe state. The curves for the 1-\textsc{%
click} and the 5-\textsc{click} outcomes have very different shapes with
small overlap: while the 1-\textsc{click} exhibits a well defined peak for
low $|\alpha |^{2}$, the 5-\textsc{click} peak is much broader and has a
maximum at higher $|\alpha |^{2}$. This shows that different measurement
outcomes are associated with easily distinguishable responses to a spanning
set of input $\alpha ,$ suggesting that the mathematical inversion from the
estimated probabilities (e.g., with Eq.\ (\ref{trace})) is practical.

An analogous tomography device exists in the area of Homodyne State Tomography: the eight-port homodyne system. This device projects the unknown state onto a coherent state, much the same as our proposed testbed. It has been pointed out that reconstruction of the density matrix from the measured data in an eight-port homodyne system is an ill-conditioned problem \cite{Leonhardt1997}. Noise in the measured data causes the inversion of Eq.\ (\ref{sumtoone}) to have large errors, indicating that while coherent states form a tomographically complete basis, they are still deficient for tomography. Considering noise from counting statistics, in detector tomography the data set size is limited only by the repetition rate of the laser, which can be as high as 80 MHz. Thus, we can quickly accumulate enough data such that counting errors for each outcome are insignificant. However, sources of noise, other than from counting, can hamper inversion. To counter the effect of these a common strategy is to introduce some amount of regularization to the inversion \cite{Boyd2004}. Often, this regularization is implicit in the reconstruction, as in the Radon Transformation or Pattern Functions used in two-port homodyne tomography, which introduce smoothing of the data \cite{Raymer2004}. Indeed, filtering of data is a common technique established in the origins of tomography in medical computer imaging. These techniques have yet to be used in detector tomography, but we expect their application will be fruitful.

\begin{widetext}

\section{A phase-space representation of the POVM}

A common complaint about quantum tomography is that the end result is
difficult to interpret physically \cite{O'Brien2004, Lanyon2007, Rhode2005,
Mitchell2003}. For example, in process tomography the reconstructed map for
the process has $d^{2}\times d^{2}$ elements for a $d$ dimensional Hilbert
space and thus quickly becomes too large to understand upon inspection. A
reconstructed POVM set can similarly contain a large number of independent
parameters, $(D-1)\times d^{2}$ if there are $D$ measurement outcomes.
However, the measurement operators $\{\hat{O}_{\gamma }\}$ benefit from
their mathematical similarities to density states, allowing us to apply many
of the same representation methods. Our detectors operate in the Fock space, which can be represented in the $x$ and $p$ basis,
where $x$ and $p$ are the normalized electric field quadratures. This points
to a particularly appealing representation commonly used to represent the
state of a field mode, the Wigner function.

\begin{figure}[h]
\centering{\ \includegraphics[width=0.7\textwidth]{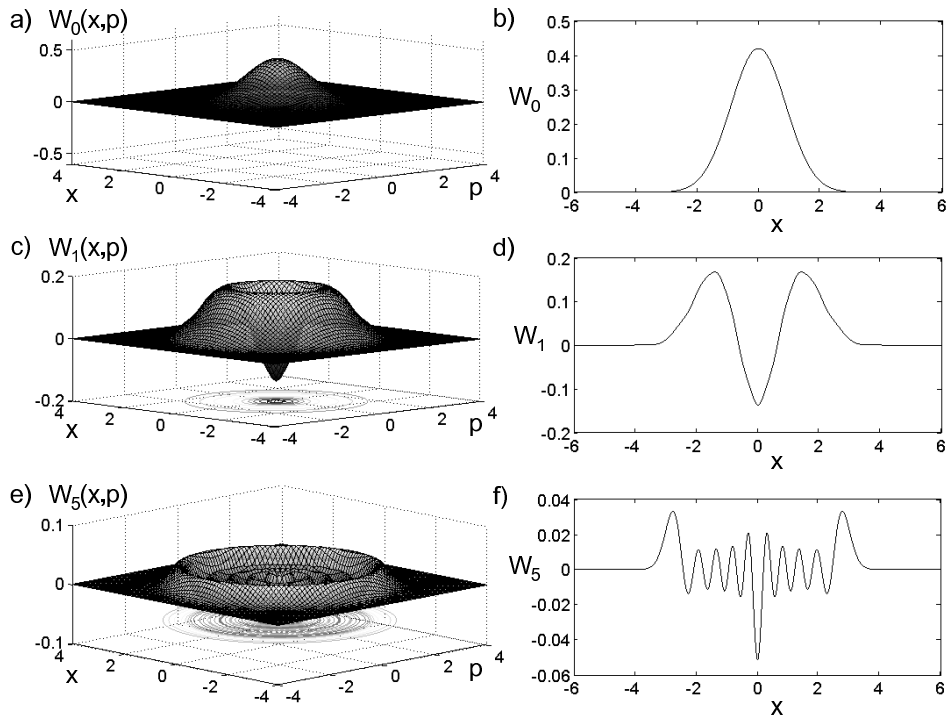} }
\caption{Wigner function representations of the a) 0-\textsc{click}, c) 1-%
\textsc{click}, and e)\ 5-\textsc{clicks} measurement outcomes of our
eight-bin time-multiplexed detector. A cross-section of each Wigner function
is shown in b), d), and f), respectively. The Wigner functions are derived
from a theoretical model of the POVMs of our eight-bin time-multiplexed
(TMD) detector.}
\label{Wigner}
\end{figure}

\end{widetext}
The Wigner function $W_{\gamma}$ is calculated in the standard way from the
POVM element $\hat{O}_{\gamma }$ \cite{Wigner1932}:

\begin{equation}
W_{\gamma }(x,p)=\frac{1}{\pi \hbar }\int_{-\infty }^{\infty }dy\,\langle
x-y|\hat{O}_{\gamma }|x+y\rangle e^{2ipy/\hbar }.
\end{equation}%
Since the measurement operators $\{\hat{O}_{\gamma }\}$ 
do not have unit trace,
this detector Wigner function is not normalized, 
\begin{equation}
\int_{-\infty }^{\infty }dx\,\int_{-\infty }^{\infty }dpW_{\gamma }(x,p)<1,
\end{equation}%
and its marginals should not be interpreted as probability distributions.
Nonetheless, it retains some appealing features. For example, the
probability of measurement outcome $\gamma $ is,
\begin{equation}
p_{\gamma }=Tr(\hat{\rho}\hat{O}_{\gamma })=\int_{-\infty }^{\infty
}dx\,\int_{-\infty }^{\infty }dpW(x,p)W_{\gamma }(x,p),~
\end{equation}%
where~$W$ is the standard Wigner function of the input state $\hat{%
\rho}.$ Thus one can visualize the measurement as the overlap of the two
Wigner functions.

We plot the Wigner functions of the theoretical measurement operators of the
TMD in Figure \ref{Wigner}. The TMD has no phase sensitivity and so the
Wigner function for each measurement operator is rotationally symmetric
around a vertical axis through the origin. Thus, on the right of Figure \ref%
{Wigner} we also present a cross-section of each Wigner function to show its
form more clearly. The Wigner functions are found from the TMD POVM\ that
was derived earlier, assuming a loss of 48\%. We show three of the nine
measurement operators, namely 0-\textsc{click}, 1-\textsc{click} and 5-%
\textsc{click} (in the figure, these are $W_{0}(x,p),$ $W_{1}(x,p)$ and, $%
W_{5}(x,p)$, respectively). Remarkably, the 0-\textsc{click} and 1-\textsc{%
click} Wigner functions look very similar to their state counterparts, the
vacuum state and single-photon state. This is despite the fact that there
are contributions from all incoming photon numbers to both (i.e., four
incoming photons might all be lost and thus result in a 0-\textsc{click}
event, or two incoming photons might end up in the same time bin and thus
result in 1-\textsc{click}). The 0-\textsc{click} Wigner function has a
Gaussian profile similar to the vacuum Wigner function, whereas the 1-%
\textsc{click} event goes negative at the origin just as the single-photon
state does.

The Wigner functions for higher click numbers quickly diverge from their
photonic equivalents though. The 5-\textsc{click} Wigner function is
significantly different from the five photon Wigner function; the former has
nine radial nodes, whereas the latter has five (the number of nodes equals
the photon number for Fock states). Since there are only eight bins, there
is a sizeable probability that six or seven incoming photons entered only
five bins in total. Consequently, there will be significant contributions to
the 5-\textsc{click} Wigner function from these higher photon numbers,
distorting it from the ideal five photon Wigner function. In contrast, the
probability that two or three incoming photons entered only one bin is
relatively small, which explains why 1-\textsc{click} Wigner function
contains much less distortion.

\section{Conclusion}

We have detailed a proposal for performing detector tomography on two detectors operating in the Fock space.
The proposed testbed generates a sequence of calibrated weak coherent states against
which to reference a detector. We have shown that these coherent states form
a tomographically complete set for the two detectors, and thus, a
tomographic reconstruction of all the POVM elements should be possible. In
the near future, we plan to perform detector tomography on these two
detectors. However, we envision that a reconstruction of the POVM elements
in the high-dimensional photon number space will present substantial
computational challenges. The reconstruction will also have to deal with
noise and uncertainty in the input states. This is usually not considered in
state tomography, where the measurements are considered to be close to ideal.

With the demonstration of detector tomography, experimentalists will have
general procedures for characterizing all the parts of a general quantum
device: the input states, the quantum circuit, and, now, the measurement. We
expect this detector tomography to be particularly useful for devices that
depend on generalized measurements for optimal functioning, such as state or
process discriminators (i.e., unambiguous state discrimination \cite%
{Peres1988,Mohseni2004}). Detector tomography should also be useful for
devices where measurement drives logic such as in cluster-state computing or
linear optics quantum computing.

\section*{Acknowledgements}

This work has been supported by the European Commission under the Integrating
Project Qubit Applications (QAP) 
and the Strep COMPAS, 
the EPSRC grant  
EP/C546237/1, and by the QIP-IRC project.
HCR has been supported by the European Commission 
under the Marie Curie Program and by the Heinz-Durr 
Stipendienprogamm of the
Studienstiftung des deutschen Volkes. 
MBP holds a Royal Society Wolfson
Research Merit Award, JE a EURYI Award.


\label{lastpage}
\end{document}